\documentclass[twocolumn, prl, aps, superscriptaddress, longbibliography,showpacs,amsmath,amssymb,floatfix
]{revtex4-2}
\usepackage{changes}
\usepackage{graphicx}
\usepackage{dcolumn}
\usepackage{bm}
\usepackage{graphicx}                                   
\usepackage{amssymb}

\usepackage{amsmath}
\usepackage{epsfig}
\usepackage{xcolor}
\usepackage{tabu}
\usepackage{mathtools}
\usepackage[colorlinks,linkcolor=blue,anchorcolor=blue,citecolor=blue,urlcolor=blue]{hyperref}
\usepackage{physics}
\usepackage{float}
\usepackage{diagbox}
\usepackage{inputenc}

\begin{document}
\title{Skin mode tunability and self-healing effect in photonic Floquet lattices}
\author{Hua-Yu Bai}
\affiliation{Laboratory of Quantum Information, University of Science and Technology of China, Hefei 230026, China}
\affiliation{Anhui Province Key Laboratory of Quantum Network, University of Science and Technology of China, Hefei, Anhui, 230026, China}
\affiliation{CAS Center For Excellence in Quantum Information and Quantum Physics, University of Science and Technology of China, Hefei 230026, China}
\affiliation{Hefei National Laboratory, University of Science and Technology of China, Hefei, Anhui, 230088, China}

\author{Yang Chen}
\email{chenyang@ustc.edu.cn}
\affiliation{Laboratory of Quantum Information, University of Science and Technology of China, Hefei 230026, China}
\affiliation{Anhui Province Key Laboratory of Quantum Network, University of Science and Technology of China, Hefei, Anhui, 230026, China}
\affiliation{CAS Center For Excellence in Quantum Information and Quantum Physics, University of Science and Technology of China, Hefei 230026, China}
\affiliation{Hefei National Laboratory, University of Science and Technology of China, Hefei, Anhui, 230088, China}

\author{Tian-Yang Zhang}
\affiliation{Laboratory of Quantum Information, University of Science and Technology of China, Hefei 230026, China}
\affiliation{Anhui Province Key Laboratory of Quantum Network, University of Science and Technology of China, Hefei, Anhui, 230026, China}
\affiliation{CAS Center For Excellence in Quantum Information and Quantum Physics, University of Science and Technology of China, Hefei 230026, China}
\affiliation{Hefei National Laboratory, University of Science and Technology of China, Hefei, Anhui, 230088, China}

\author{Guang-Can Guo}
\affiliation{Laboratory of Quantum Information, University of Science and Technology of China, Hefei 230026, China}
\affiliation{Anhui Province Key Laboratory of Quantum Network, University of Science and Technology of China, Hefei, Anhui, 230026, China}
\affiliation{CAS Center For Excellence in Quantum Information and Quantum Physics, University of Science and Technology of China, Hefei 230026, China}
\affiliation{Hefei National Laboratory, University of Science and Technology of China, Hefei, Anhui, 230088, China}

\author{Ming Gong}
\email{gongm@ustc.edu.cn}
\affiliation{Laboratory of Quantum Information, University of Science and Technology of China, Hefei 230026, China}
\affiliation{Anhui Province Key Laboratory of Quantum Network, University of Science and Technology of China, Hefei, Anhui, 230026, China}
\affiliation{CAS Center For Excellence in Quantum Information and Quantum Physics, University of Science and Technology of China, Hefei 230026, China}
\affiliation{Hefei National Laboratory, University of Science and Technology of China, Hefei, Anhui, 230088, China}

\author{Xi-Feng Ren}%
\email{renxf@ustc.edu.cn}
\affiliation{Laboratory of Quantum Information, University of Science and Technology of China, Hefei 230026, China}
\affiliation{Anhui Province Key Laboratory of Quantum Network, University of Science and Technology of China, Hefei, Anhui, 230026, China}
\affiliation{CAS Center For Excellence in Quantum Information and Quantum Physics, University of Science and Technology of China, Hefei 230026, China}
\affiliation{Hefei National Laboratory, University of Science and Technology of China, Hefei, Anhui, 230088, China}

\begin{abstract}
Non-Hermitian systems host exotic phenomena absent in their Hermitian counterparts, including the recently predicted self-healing effect (SHE) of non-Hermitian skin modes. To date, the SHE of skin modes in non-Hermitian systems has not been observed experimentally. Here we propose a feasible scheme to realize SHE in photonic Floquet lattices by exploiting skin mode tunability (SMT), a mechanism in which the spectrum of  skin modes localized at one boundary can be tuned via a potential applied at the opposite boundary. Such tunability arises from the non-Hermitian biorthogonality of the eigenstates. We demonstrate that a certain skin mode is exceptionally sensitive to remote-boundary potentials in an array of $100$ coupled helical waveguides, allowing broad-range spectral control and the generation of SHE with experimentally accessible parameters [M. C. Rechtsman \textit{et al.}, \href{https://doi.org/10.1038/nature12066}{Nature  {\bf 496}, 196–200 (2013)}; Y. Sun \textit{et al.}, \href{https://doi.org/10.1103/PhysRevLett.132.063804}{Phys. Rev. Lett. {\bf 132}, 063804 (2024)}]. Our results establish a general framework for engineering skin modes via local perturbations, thereby expanding the toolbox for non-Hermitian wave control.

\end{abstract}

\maketitle
Self-healing effect (SHE) was first demonstrated in conventional diffraction-free systems \cite{bouchal1998self}, where light fields recover their spatial profiles after encountering an obstacle because the diffracted fields decrease inversely with propagation distance $z$. Longhi later extended this concept to non-Hermitian systems, where the lattice typically exhibits the non-Hermitian skin effect (NHSE) \cite{zhang2020correspondence,yi2020non,li2022gain,yang2020non,okuma2020topological,song2019non,song2019non1,zhang2022universal,xue2022non,longhi2020non,borgnia2020non,li2020critical,lee2019hybrid,edvardsson2022sensitivity,lin2023topological}, predicting infinitely many self-healing skin modes that are exponentially localized at the lattice boundary \cite{longhi2022self}. This extension builds on key principles of non-Hermitian physics, including biorthogonal eigenstates \cite{brody2013biorthogonal, Kunst2018Biorthogonal, Ashida2020non} and non-Bloch Band theory\cite{yao2018edge,shen2018topological,gong2018topological,kawabata2019symmetry,yokomizo2019Non}. The exploration of non-Hermitian phenomena has been experimentally realized in various platforms, including photonics \cite{wang2021generating,zhu2022anomalous,weidemann2020topological,xiao2020non,lin2023manipulating,liu2022complex,zhou2023observation,sun2024photonic,lin2024observation}, electrical circuits \cite{helbig2020generalized,zou2021observation}, ultracold atoms \cite{liang2022dynamic,zhao2025two}, acoustic systems \cite{zhang2021acoustic,zhang2021observation,gao2022anomalous,gu2022transient} and mechanical lattices \cite{wang2022non,li2024observation}.

A key characteristic of non-Hermitian systems is their high spectral sensitivity to boundary conditions \cite{REICHEL1992153,gong2018topological}. The boundary modification enabling transition between periodic boundary conditions (PBC) and open boundary conditions (OBC) in non-Hermitian systems reshapes the entire spectral and spatial characteristics of skin modes \cite{Xiong2018bulk,li2021impurity,guo2021exact,liu2021exact,hu2024non}. Unlike in Hermitian lattices, where defects induce localized states around the perturbation \cite{Kosterlitz2017Nobel}, recent studies have further shown that by engineering edge terms under OBC, one can selectively excite the quasi-edge states under semi-infinite boundary conditions \cite{longhi2022self,longhi2022selective}, and realize counter skin effect \cite{leumer2025impurity}. However, in systems under OBC, the response of the energy spectrum to boundary perturbations and the feasible experimental routes for achieving SHE remain unclear.

Here, we propose and demonstrate skin mode tunability (SMT), a mechanism for tuning the spectrum of skin modes by applying boundary modifications at the edge opposite to their localization. While skin modes exhibit strong spatial confinement under OBC, the biorthogonal nature of non-Hermitian systems enables precise spectral tuning through remote end perturbations. By adjusting boundary potentials, a specific skin mode can be isolated from the other states, forming the self-healing state (SHS) with the largest imaginary energy that survives during propagation after scattering. This mechanism enables the realization of SHE for skin modes, and we further propose an experimentally feasible implementation using photonic Floquet lattices with currently accessible parameters.

\begin{figure}[htbp]
\includegraphics[width=0.46\textwidth]{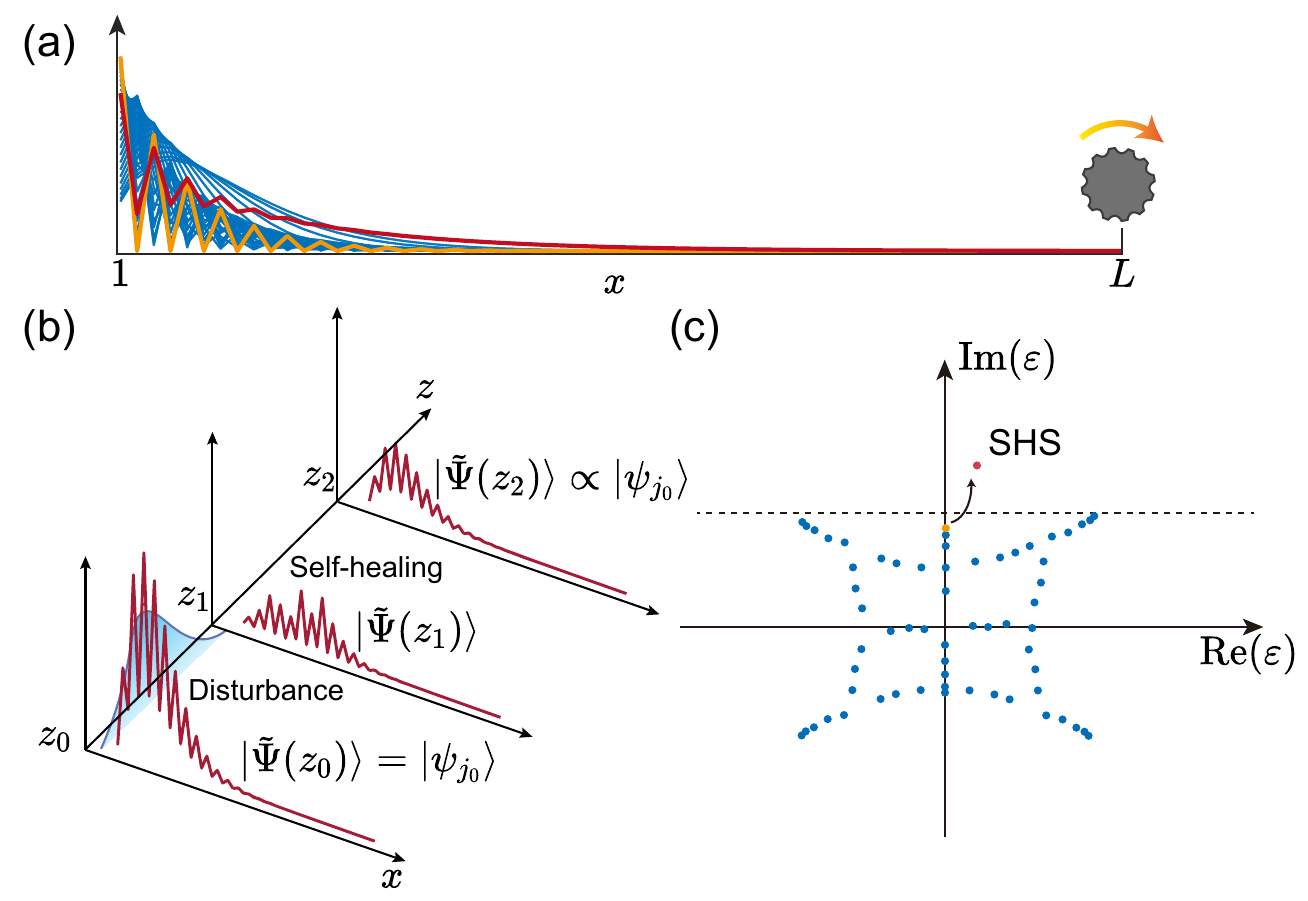}
\caption{(a) Tunability of a skin mode localized at the left boundary by a local perturbation acting on the right end. (b) Schematic of SHE, which demonstrates the wave function recovers to its original form after encountering a disturbance. (c) Mechanism of the SHE based on the SMT shown in (a), where the SHS is spectrally isolated with the largest imaginary eigenenergy, $\Im \epsilon_{j_0}$. }
\label{fig1}
\end{figure}

{\it General idea for SMT and SHE}.---The general idea of SMT in non-Hermitian systems exhibiting NHSE can be introduced by analyzing a simple model under OBC: $H = p^2/2 + i\alpha p$, where $\alpha$ represents the strength of dissipation ($\alpha<0$) or gain ($\alpha>0$). The eigenstates of $H$ are given by $|\psi_j^R\rangle=c_R \mathrm{ sin}\frac{j\pi x}{L} e^{\alpha x}$, and those of  $H^\dagger$ are $
|\psi_j^L\rangle=c_L \mathrm{ sin}\frac{j\pi x}{L} e^{-\alpha x}$. Here,  $|\psi_j^{R(L)}\rangle$ are the right (left) eigenstates labeled by $j$, and the condition $c_L^\ast c_R=\frac{2}{L} $ ensures biorthogonality \cite{brody2013biorthogonal, Ashida2020non}. For a perturbation $\delta V = \delta v \delta(x-x_0)$, the SMT is defined as
\begin{equation}
    \chi_{j}={\langle \psi_j^L |\delta V |\psi_j^R \rangle \over \delta v} =\frac{2}{L} \sin ^{2}\left(\frac{j \pi x_{0}}{L} \right),
    \label{eq-chidef}
\end{equation}

which is a measure of the eigenenergy response to the change in the potential at the lattice end $x_0$. Notably, $\chi_j$ is independent of $\alpha$, demonstrating that skin modes in an open system, despite their vanishing amplitude at one boundary, retain energy sensitivity to edge perturbations.

This tunability stems from a fundamental argument. Consider a sufficiently large $L$-site chain where skin modes are strongly localized on the left end, exhibiting negligible amplitude at the right end [Fig. \ref{fig1}(a)]. If we remove the rightmost site, conventional perturbation theory would suggest the wave functions remain largely unchanged. However, this would imply $L$ skin modes with $L-1$ lattice sites, a contradiction. Thus, even if the amplitude of the wave function is negligible at the boundary, these states must be sensitive to local perturbations. The failure of perturbation theory arises due to the non-orthogonality of eigenstates in non-Hermitian Hamiltonians, requiring separate left and right eigenstates \cite{brody2013biorthogonal, Kunst2018Biorthogonal,Ashida2020non}. Consequently, the SMT emerges as an intrinsic property of such systems.

As a direct manifestation of the SMT, we explore its possible application in SHE. Consider a non-Hermitian Hamiltonian $H$ with right eigenstates $|\psi_j^{R}\rangle$ and left eigenstates $|\psi_j^{L}\rangle$, satisfying $H|\psi_j^{R}\rangle=\epsilon_j |\psi_j^{R}\rangle$ and $H^\dagger|\psi_j^{L}\rangle=\epsilon_j^\ast |\psi_j^{L}\rangle$. We focus on the time evolution (where $z$ denotes time) of the eigenstate $|\psi_{j_0}\rangle$, which corresponds to the eigenenergy with the largest imaginary part and let $|\Psi(z)\rangle$ be the evolved wave function in the absence of the disturbance. As shown in Fig. \ref{fig1}(b), suppose the system is subjected to a disturbance between $z_0$ and $z_1$, and the wave function after the disturbance is $|\tilde{\Psi}(z_1)\rangle=\sum_{j} c_{j} |\psi_{j}^R\rangle$, where $c_{j}=\langle\psi_{j}^{L} \mid \tilde{\Psi}(z_1)\rangle$.The subsequent dynamics for $z>z_1$ is governed by 
\begin{equation}
    |\tilde{\Psi}(z)\rangle= e^{-i H (z-z_1)}|\tilde{\Psi}(z_1)\rangle=\sum_{j} c_{j} e^{ -i\epsilon_{j} (z-z_1)} |\psi_{j}^R\rangle.
\end{equation} 
At long times, the evolution is dominated by $|\psi_{j_0}\rangle$, as all other states decay rapidly due to their smaller gain rates. The SHE is characterized by $\eta(z)=1-\left |\langle\Psi_\text{norm}(z) \mid \tilde{\Psi}_\text{norm}(z) \rangle\right |$, where $\lim_{z\to\infty} \eta(z)=0$ indicates perfect self-healing. Here, $|\Psi_\text{norm}(z)\rangle$ and $|\tilde{\Psi}_\text{norm} (z)\rangle$ are the normalized wave functions of $|\Psi(z)\rangle$ and $|\tilde{\Psi}(z)\rangle$. The self-healing property of $|\psi_{j_0}\rangle$ emerges when it is spectrally isolated [Fig. \ref{fig1}(c)], which we achieve here through edge perturbations. This idea is conceptually analogous to the imaginary-time evolution method for ground-state preparation \cite{hoffmann1996computational}, but instead isolates the skin mode with maximum gain.

\begin{figure}[t]
\includegraphics[width=0.46\textwidth]{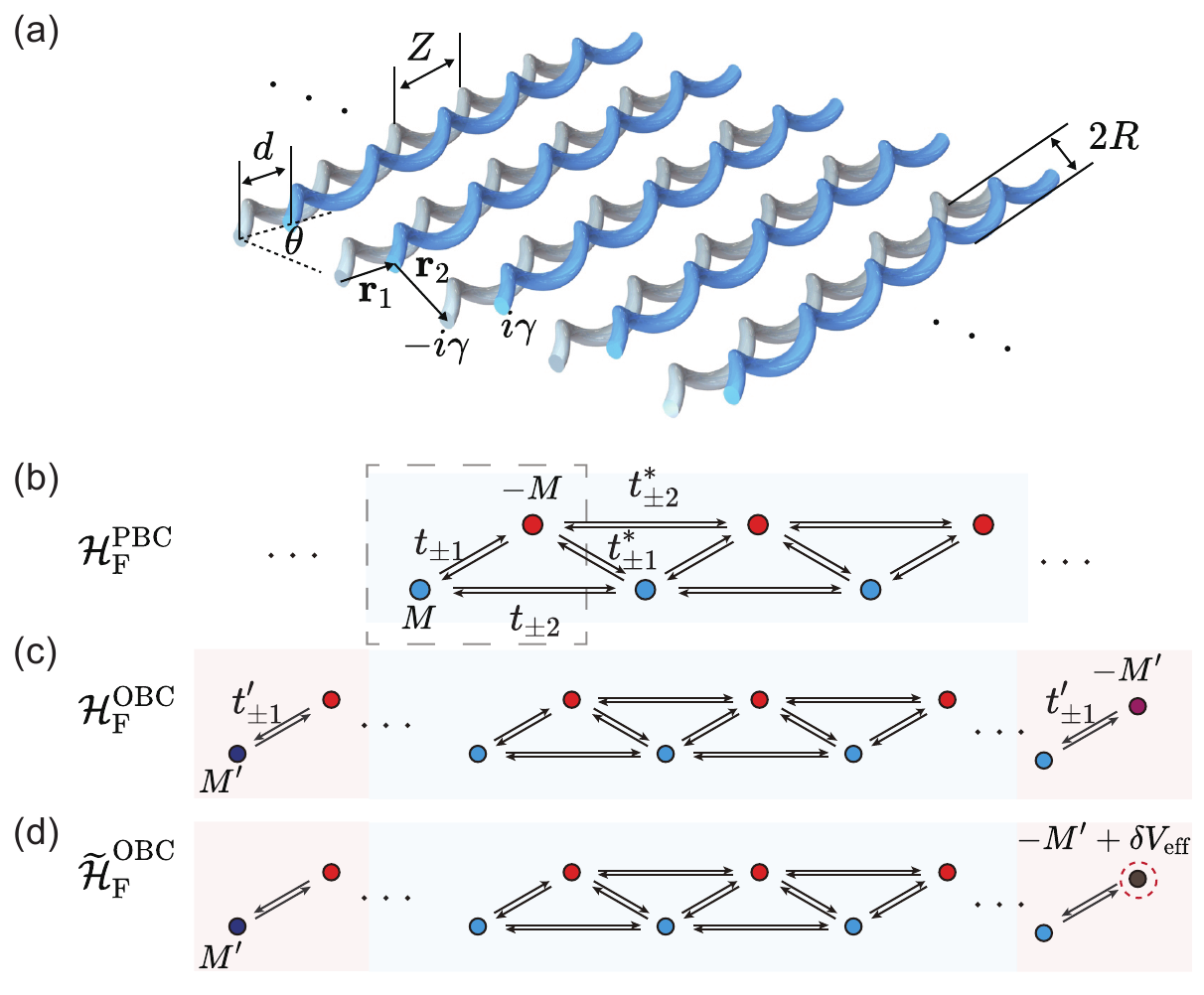}
\caption{\label{fig-epsart}(a) Floquet lattice for realizing the SHE in a zigzag helical waveguide array.  Schematic representation of (b) the effective Hamiltonian under PBC, $\mathcal{H}_\text{F}^{\text{PBC}}$, (c) the effective Hamiltonian under OBC, $\mathcal{H}_\text{F}^{\text{OBC}}$, and (d) the OBC Hamiltonian with modified boundary term $\delta V_{\rm eff}$. Long-range interactions beyond next-nearest neighbors are not shown for clarity in (b)-(d).}
\label{fig2}
\end{figure}

{\it Floquet lattice, non-Hermitian Hamiltonian and winding number}.---We now demonstrate SMT and SHE in non-Hermitian Floquet lattices. Floquet lattices realize an effective magnetic field by breaking time reversal symmetry, enabling anomalous topological insulating phases \cite{aidelsburger2011experimental,atala2014observation,rechtsman2013photonic,maczewsky2017observation,yang2020photonic}. The interplay between the synthetic gauge field and gain/loss gives rise to the NHSE \cite{yi2020non,li2022gain,lin2021steering,wu2022flux,li2023loss,sun2024photonic}. 
In our implementation, the Floquet lattices consists of helical waveguides arranged in zigzag chains [Fig. \ref{fig2}(a)], where each unit cell contains two waveguides with loss and gain of $\pm \gamma$. The light propagation dynamics is governed by Maxwell's equations, which maps to a tight-binding Hamiltonian \cite{marte1997paraxial,longhi2009quantum,rechtsman2013photonic,chen2021tight}
\begin{eqnarray}
    &&  \mathcal{H}(z) = \sum_l -i\gamma c_{2l-1}^\dagger c_{2l-1} + i\gamma c_{2l}^\dagger c_{2l} +  \label{eq-Hz} \\
       && (\kappa e^{-i\mathbf{A}(z)\cdot\hat{\mathbf{r}}_1} c_{2l-1}^\dagger c_{2l} 
       + \kappa e^{-i\mathbf{A}(z)\cdot \hat{\mathbf{r}}_2} c_{2l}^\dagger c_{2l+1} + \text{h.c.}),\nonumber
\end{eqnarray} 
where $c_l^\dagger$ $(c_l)$ is the creation (annihilation) operator at site $l$; $z$ is the propagation direction. The gauge potential is $\boldsymbol{A}(z)=k_{0} R \Omega d(\sin (\Omega z), -\cos (\Omega z), 0)$, where $k_0=2 \pi n_{0} / \lambda$ is the wavenumber (with $n_0$ being the refractive index of the substrate and $\lambda$ the wavelength), $R$ is the helical radius, $\Omega=2\pi/Z$ is the frequency of rotation (where $Z$ donates t he period), and $d$ is the nearest-neighbor waveguide distance. The nearest-neighbor coupling coefficient $\kappa$ depends on $d$. The vectors $\hat{\mathbf{r}}_{1,2} = (\cos\theta,\pm \sin\theta)$ characterize the zigzag structure with an angle $\theta$. 

The effective Hamiltonian $\mathcal{H}_\text{F}$ of a Floquet system under different boundary conditions is derived from the time evolution operator over a full period $Z$ \cite{lindner2011floquet,jotzu2014experimental,dalibard2011colloquium,goldman2014periodically,bandres2016topological}. For our system, we numerically compute $\mathcal{H}_\text{F}$ via 
\begin{equation}
\mathcal{H}_\text{F} = {i \over Z} \ln \prod_{i=1}^{K} \exp(-i\mathcal{H}(z_i) \delta z),
\label{eq-Heffdis}
\end{equation}
where $K = 1000$ represents the number of discrete steps in one period $Z$, with each step given by $\delta z = Z/K$. The resulting $\mathcal{H}_\text{F}$ is a dense non-Hermitian matrix. The matrix elements $(\mathcal{H}_\text{F})_{ll^\prime}$ are negligible when $|l-l^\prime| > 3$ \cite{li2023loss}. Unde PBC, the bulk Hamiltonian $\mathcal{H}_\text{F}$ takes the form [Fig. \ref{fig2}(b)]
\begin{equation}
\begin{split}
    \mathcal{H}_\text{F}= \sum_{\mathbf{q} = 0,\pm 1, \pm 2} \sum_{l} 
      \mathbf{\psi}_{l}^{\dagger} \mathbf{J}_{-\mathbf{q}} \mathbf{\psi}_{l+\mathbf{q}}
\end{split},
\label{eq-Hinfinity}
\end{equation}
where $\mathbf{\psi}_{l}= \left(\begin{array}{cc}
a_{l} \\
b_{l}
\end{array}\right)$, $\mathbf{J_{0}}=\left(\begin{array}{cc}
M & t_{-1} \\
t_1 & -M
\end{array}\right)$, $\mathbf{J_{-1}}=\left(\begin{array}{cc}
t_{-2} & t_{-3} \\
t_{-1}^\ast & t_{-2}^\ast
\end{array}\right)$, $\mathbf{J_{1}}=\left(\begin{array}{cc}
t_{2} & t_{1}^\ast \\
t_3 & t_{2}^\ast
\end{array}\right)$, $\mathbf{J_{-2}}=\left(\begin{array}{cc}
0 & 0 \\
t_{-3}^\ast & 0
\end{array}\right)$ and $\mathbf{J_{2}}=\left(\begin{array}{cc}
0 & t_3^\ast \\
0 & 0
\end{array}\right)$. Here $t_{\pm m}$ is the hopping between sites with distance $m$, with $\pm$ denotes the direction, and $\pm M$ is the effective onsite potential in each unit cell. For the SMT discussed in this work, the eigenvalues are exactly determined by considering boundary effects on the effective Hamiltonian under OBC [Fig. \ref{fig2}(c)], rather than averaging the coupling strength as in Ref. \cite{li2023loss}.

Following experimental parameters from Refs. \cite{rechtsman2013photonic,stutzer2018photonic,yang2020photonic,wang2022edge,sun2024photonic}, we consider waveguides with $\lambda=633 ~\mathrm{nm}$, $n_0=1.505$, and $d=11 ~\mu \mathrm{m}$ (corresponding to a coupling coefficient $\kappa=0.25 ~\mathrm{mm}^{-1}$). The helical period is $Z=5 ~\mathrm{mm}$ ensuring low bending losses, while the zigzag angle $\theta=\pi/6$ restricts $\mathcal{H}(z)$ to nearest-neighbor coupling. The setups are similar to those reported in previous works on helical waveguides \cite{rechtsman2013photonic,sun2024photonic}. The gain/loss amplitude is set to $\gamma=0.14 \kappa$, experimentally achievable as pure loss through a globally tunable decay factor, with precise control enabled by sectioned waveguide fabrication \cite{xia2021nonlinear}. 

\begin{figure}[t]
\includegraphics[width=0.48\textwidth]{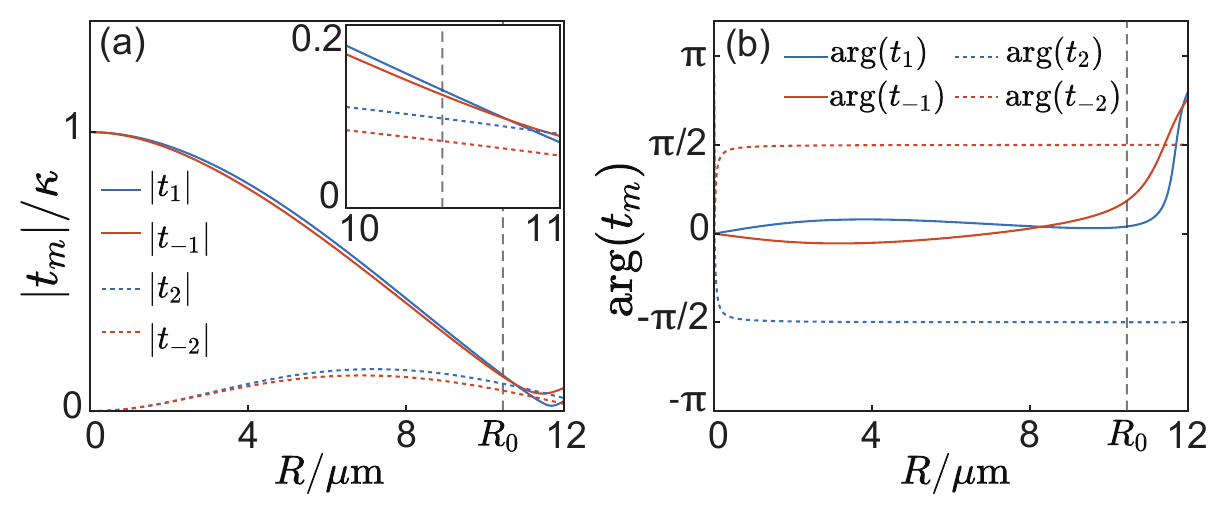}
\caption{Amplitude (a) and phase (b) of the hopping term $t_{\pm 1,\pm 2}$ in the helical waveguides as a function of $R$. We choose $R_0=10.45 ~\mu \mathrm{m}$, where the first- and second-order coupling strengths are comparable, with $t_1/\kappa=0.128 + 0.016i $, $t_{-1}/\kappa=0.104 +0.068i $, $t_2/\kappa=-0.098i $, $t_{-2}/\kappa=0.074i $. }
\label{fig3}
\end{figure}

The numerical results of the effective couplings $t_{\pm m}$  as a function of $R$ are shown in Fig. \ref{fig3}. At $R=0$, only $t_{\pm 1}=\kappa$ are present, while higher-order couplings $| m | \geq 2$ become significant as $R$ increases. The first- and second-order effective couplings in the two directions differ slightly \cite{huang2021loss,li2023loss}. Notably, $t_{\pm2}$ approximately maintain fixed phases of $\pm \pi/2$ in a certain range of helical radius. At our chosen $R=10.45 ~\mu \mathrm{m}$ (where first- and second-order couplings are comparable), we include up to third-order terms [Eq.(\ref{eq-Hinfinity})]. The dimensionless gauge field strength, given by $|A|=k_0 R \Omega d=2.16$, is comparable to the value reported in Ref. \cite{stutzer2018photonic}. 

\begin{figure}[t]
\includegraphics[width=0.48\textwidth]{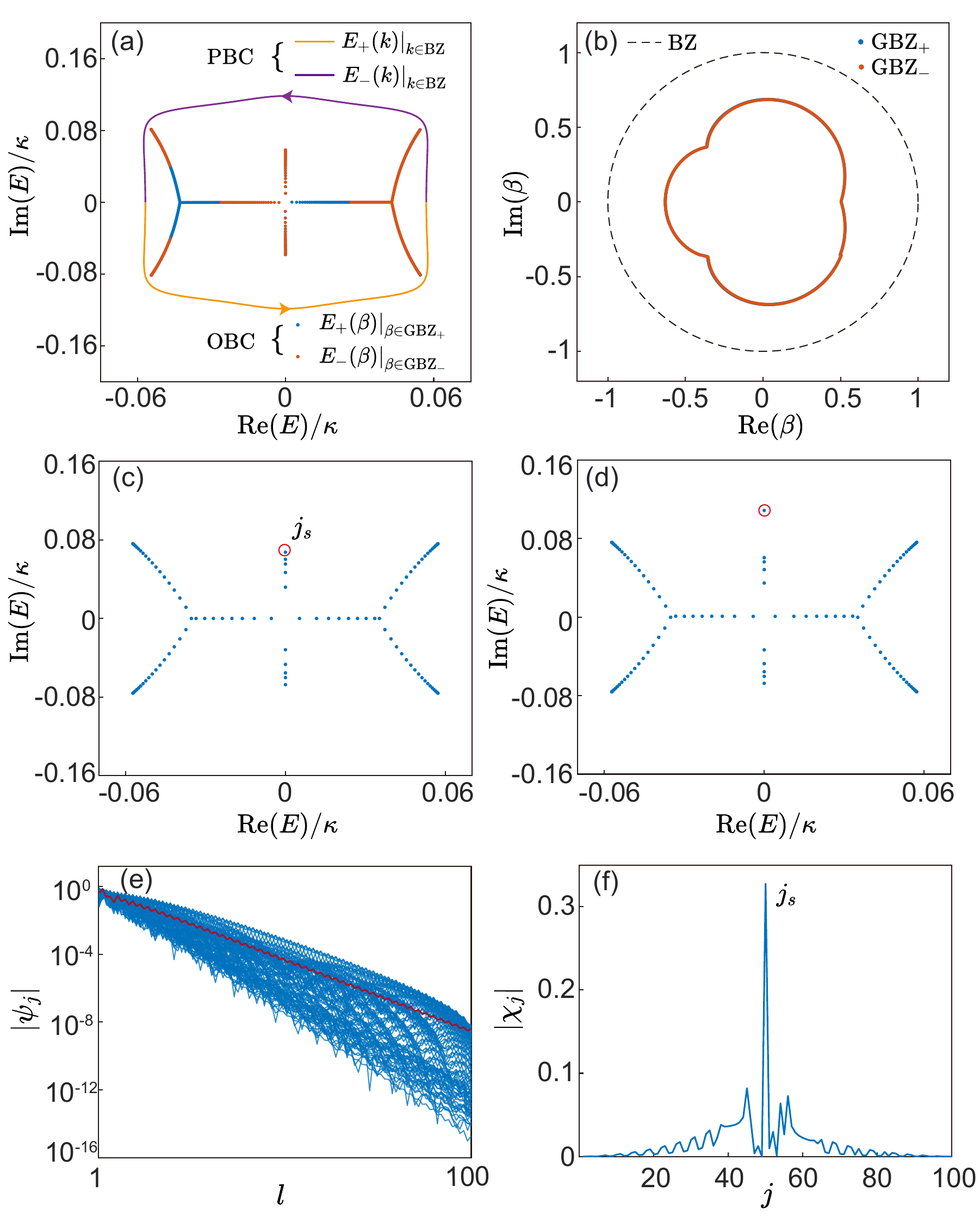}
\caption{(a) Energy spectrum of $\mathcal{H}_\text{F}^{\text{PBC}}$, which forms a loop and exhibits a nonzero winding number $W=1$. The spectrum of $\mathcal{H}_\text{F}^{\text{OBC}}$ in the thermodynamic limit is inside the spectrum of 
$\mathcal{H}_\text{F}^{\text{PBC}}$. (b) The calculated $\text{GBZ}$, which lies inside the unit circle and the two sub-GBZs nearly coincide. (c) Spectrum of $\mathcal{H}_\text{F}^{\text{OBC}}$ for $L=50$ unit cells. The state $|\psi_{j_s}\rangle$ with the maximum SMT is highlighted with a red circle. (d) Spectrum of the effective Hamiltonian $\widetilde{\mathcal{H}}_\text{F}^{\text{OBC}}$ when a controlled potential  $\delta V=0.06i\kappa$ is applied to the right end of $\mathcal{H}(z)$ under OBC. Compared with $\mathcal{H}_\text{F}^{\text {OBC}}$, the eigenvalues of most skin modes remain nearly unchanged, except for $|\psi_{j_s}\rangle$ (red circle). (e) Spatial distribution of all eigenstates $|\psi_j\rangle$, where $|\psi_{j_s}\rangle$ is highlighted in red. (f) $|\chi_j|$  of all states,  showing that $j_s=50$ corresponds to the skin mode with the maximum SMT.}\label{fig4}
\end{figure}

We investigate the NHSE using the momentum-space Hamiltonian of Eq.(\ref{eq-Hinfinity}), given by
\begin{eqnarray}
    && \mathcal{H}_\text{F}(k) =\\
&& \begin{pmatrix}
M + t_{2} /\beta + t_{-2}\beta & 
 t_{-1} + t_{1}^\ast/ \beta + t_{-3}\beta + t_{3}^\ast/\beta^2\\[1em]
 t_{1} + t_{-1}^\ast\beta + t_{3} /\beta + t_{-3}^\ast\beta^2& 
-M + t_{2}^\ast/\beta + t_{-2}^\ast\beta
\end{pmatrix},\nonumber
\end{eqnarray} 
where $\beta=e^{ik}$, and the eigenvalues of this two-band system are $E_\pm(k)$. The spectrum of  $\mathcal{H}_\text{F}^{\text{PBC}}$ forms a closed loop in the complex energy plane [Fig. \ref{fig4}(a)], where $k$ traverses the Brillouin zone (BZ) $[-\pi, \pi]$. The spectrum of $\mathcal{H}_\text{F}^{\text{OBC}}$ is obtained by removing periodic connections. In the thermodynamic limit ($L\to\infty$), its bulk spectrum is determined by $E_\pm(\beta)$, where $\beta$ varies over the generalized Brillouin zone (GBZ). As shown in Fig. \ref{fig4}(b), the sub-GBZs $\text{GBZ}_{\pm}$ lie inside the unit circle. This bulk spectrum remains unchanged under edge perturbations \cite{yao2018edge} and is entirely contained within the PBC spectrum [Fig. \ref{fig4}(a)]. The point-gap topology of $\mathcal{H}_\text{F}^{\text{PBC}}$ is characterized by the winding number:
\begin{equation}
    W(E)=\frac{1}{2 \pi i} \oint_{\mathrm{BZ}} \partial_{k} \ln \operatorname{det}\left[\mathcal{H}_\text{F}(k)-E\right] d k,
\end{equation}
which yields $W=1$ for any energy $E$ in the OBC bulk spectrum. This quantifies the spectral winding and predicts the left-boundary localization of OBC eigenstates \cite{okuma2020topological,zhang2020correspondence}.

{\it Boundary manipulation and SHE}.---Building on these results, we now present the central idea of this work: the manipulation of skin modes and the realization of SHE. Our experimental scheme is based on a Floquet system with open boundaries, consisting of $L=50$ unit cells. Under OBC, the boundary terms in Eq. \ref{eq-Hinfinity} are slightly modified due to boundary effects, as described by the Magnus expansion \cite{blanes2009magnus,goldman2014periodically,jotzu2014experimental,eckardt2015high}. Numerical calculations reveal that the boundary term takes the form
\begin{equation}
   M^{\prime} (a^\dagger_1 a_1 - b^\dagger_{L} b_{L}) + t_1^{\prime} (b^\dagger_1 a_1 + b^\dagger_{L} b_{L})+ t_{-1}^{\prime} (a^\dagger_1 b_1 +a^\dagger_L b_{L}),
\end{equation}
while the bulk terms remain approximately $t_{\pm m}$ and $\pm M$ as depicted in Fig. \ref{fig2}(c) \footnote{$M/\kappa=-0.108i $, $M^{\prime}/\kappa=0.046 - 0.123i $, $t_1/\kappa=0.128 + 0.016i $, $t_1^{\prime}/\kappa=0.121 + 0.036i $, $t_{-1}/\kappa=0.104 +0.068i $, $t_{-1}^{\prime}/\kappa=0.092 + 0.056i$}. This suggests that while the effective Hamiltonians of Floquet systems under PBC and OBC share fundamentally similar structures, realistic Floquet systems require careful consideration of boundary effects. Moreover, boundary interactions can be selectively modified without perturbing the bulk terms.

We introduce an end potential $\delta V$ on the rightmost helical waveguide in Eq. \ref{eq-Hz}. Using Eq. \ref{eq-Heffdis}, the modified effective Hamiltonian becomes
\begin{equation}
\widetilde{\mathcal{H}}_\text{F}^{\text{OBC}}\approx \mathcal{H}_\text{F}^{\text {OBC}}+\delta V_{\text{eff}} b_{L}^\dagger b_{L},
\end{equation}
where $\delta V_{\text{eff}} \sim \delta V$ [Fig. \ref{fig2}(d)]. Since $\delta V$ is a small, time-independent perturbation, it does not significantly alter other terms.

\begin{figure}[t]
\includegraphics[width=0.48\textwidth]{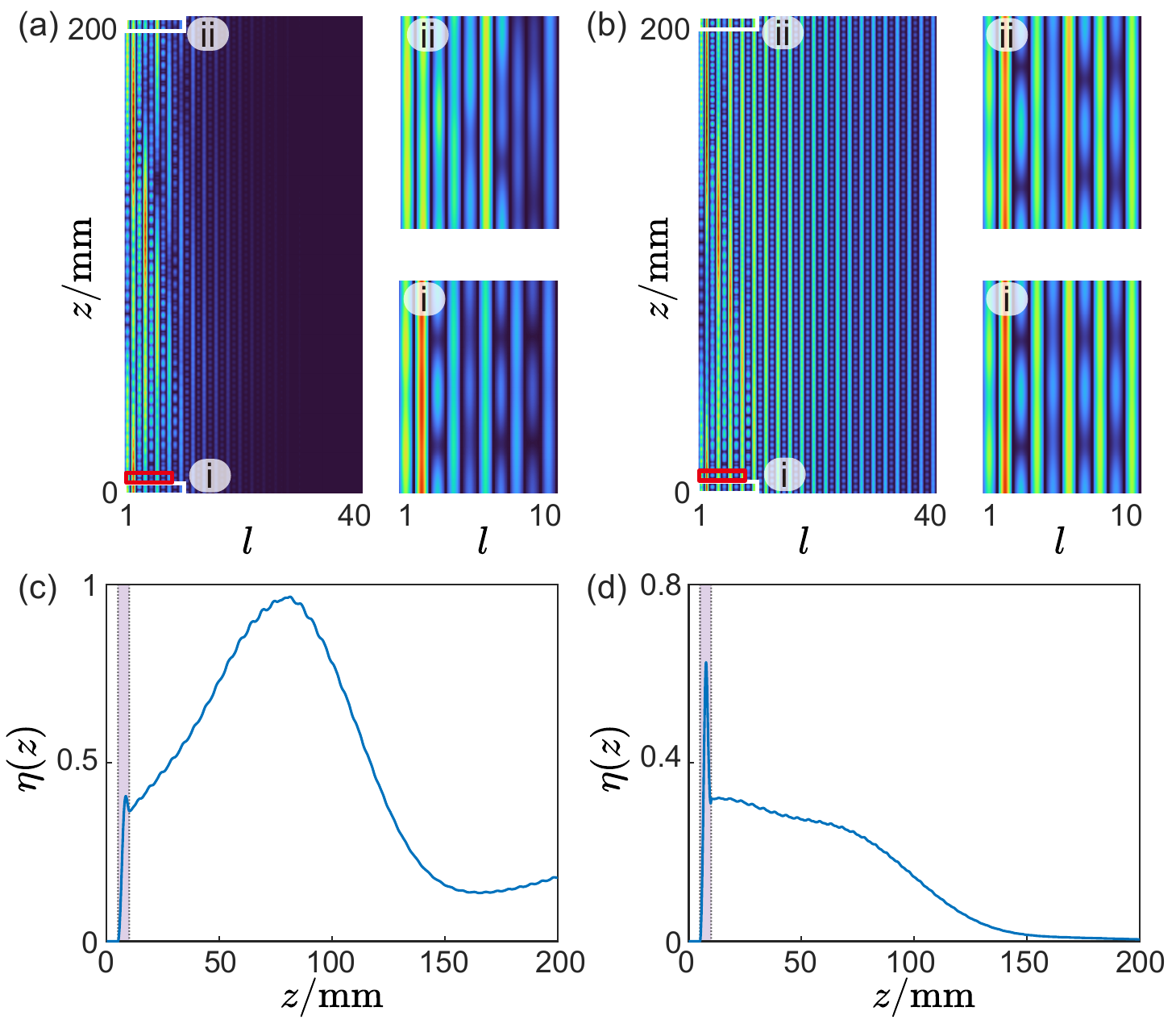}
\caption{SHE of a skin mode via SMT in a Floquet waveguide lattice. (a) Numerical intensity distribution along the propagation direction before adding the end potential perturbation, with input state $|\psi_{j_s}\rangle$, which is not spectrally isolated ($E/\kappa =0.068i$). The result shows that the state does not exhibit self-healing. (b) Propagated intensity distribution after applying the edge perturbation, with the input state $|\psi_{j_s}\rangle$ now being an SHS ($E/\kappa =0.109i$). In both (a) and (b), disturbance in the helical waveguides is indicated by the red rectangle, and only the first 40 waveguides are shown.  (c),(d) Deviation of the evolved wavefunction with disturbance, $|\tilde{\Psi}_\text{norm} (z)\rangle$, from that without disturbance, $|\Psi_\text{norm}(z)\rangle$, for (a),(b), defined as $\eta(z)=1-\left |\langle\Psi_\text{norm}(z) \mid \tilde{\Psi}_\text{norm}(z) \rangle\right |$. The region marked in purple indicates the interval during which the disturbance occurs.}
\label{fig5}
\end{figure}

The spectrum of $\mathcal{H}_\text{F}^{\text {OBC}}$ for $L=50$ unit cells without the end perturbation is shown in Fig. \ref{fig4}(c). The spatial distribution of eigenstates $|\psi_j\rangle$ [Fig. \ref{fig4}(e)] reveals that all wave functions decay to zero at the right end, which is a signature of the NHSE in the Floquet lattice. We calculate the SMT $\chi_j$ as shown in Fig. \ref{fig4}(f), where states are indexed in ascending order of the real part of energy. The tunable eigenstate is identified as the state with the largest SMT ($j_s=50$). As shown in Fig. \ref{fig4}(d), upon introducing a potential $\delta V = 0.06i \kappa$, the state exhibiting the strongest response corresponds to the maximum SMT, rather than to the state with the largest amplitude at the modified boundary. As $\delta V$ increases, this state becomes spectrally isolated, confirming our SMT prediction. The self-healing condition is met by tuning $\delta V$ to control the energy of the skin mode, where the gain and loss can be modulated in experiments.

Finally, we discuss the SHE in the Floquet lattices with a modulated boundary potential as shown in Fig. \ref{fig5}. In our numerical simulations, we introduce a Gaussian refractive index deviation $\Delta n_\text{eff}(l)=10^{-4}e^{-0.025 l^2}$ in the helical waveguides with $l\le8$ within $z \in [Z, 2Z]$ as a disturbance of the wave function. The refractive index deviation can be experimentally implemented via femtosecond laser direct writing \cite{corrielli2013fractional,chen2025reentrant}. The initial input state is the state with the maximum $\chi$, i.e., $|\psi_{j_s}\rangle$ [Fig. \ref{fig4}(c), (d)]. Without boundary perturbation ($\delta V=0$), the input state $|\psi_{j_s}\rangle$ fails to self-heal as shown in Fig. \ref{fig5}(a) and (c). However, when a boundary perturbation $\delta V=0.06i\kappa$ is applied at the right edge, $|\psi_{j_s}\rangle$ becomes an SHS with $\eta(z) \to 0$ [Fig. \ref{fig5}(b), (d)]. These perturbations selectively generate SHS while maintaining minimal impact on other states, demonstrating the unique self-healing capability of the engineered skin mode via edge perturbations.

{\it Conclusion and Discussion}.---We introduce the concept of SMT, which enables the control of the spectrum of non-Hermitian skin modes by applying a potential at the boundary opposite to the localized side. This effect, with no Hermitian analogue, enables the realization of SHE of skin modes. We propose a concrete experimental implementation of our results in photonic helical waveguides. Even in passive systems without gain, global dissipation combined with local boundary modulation can suppress the loss of the SHS to levels that permit clear experimental observation of SHE. Moreover, SMT provides a controllable platform to probe the interaction between skin modes and local perturbations at arbitrary positions, offering insights into scaling behavior in critical regimes \cite{li2020critical,guo2021exact,liu2021exact}. This work proposes a framework for steering the energy and localization of skin modes via local perturbations, extending the control paradigm of non-Hermitian wave physics \cite{feng2017non,han2019non}. The state most sensitive to the end perturbation remains localized at the left boundary under our parameters, but it is expected to shift to the right boundary under sufficiently strong perturbations. Investigating this transition and its connection to the topological origin of the skin effect remains an open question.

\textit{Acknowledgments}: This work is supported by the National Key R\&D Program of China (No. 2022YFA1204704), the National Natural Science Foundation of China (NSFC) (Nos. T2325022, U23A2074, 12204462), the Strategic Priority Research Program of the Chinese Academy of Sciences (Grant No. XDB0500000), CAS Project for Young Scientists in Basic Research (No.253 YSBR-049), the Innovation Program for Quantum Science and Technology (2021ZD0303200, 2021ZD0301200, 2021ZD0301500), and Fundamental Research Funds for the Central Universities (WK2030000107, WK2030000108). This work was partially carried out at the USTC Center for Micro and Nanoscale Research and Fabrication.

H. Y. Bai and Y. Chen contributed equally to this work.

\bibliography{ref}

\end{document}